\documentclass[amssymb,onecolumn,aps, a2paper, 11pt]{revtex4}

\usepackage{psfrag}
\usepackage{amsmath}
\usepackage{amsfonts}
\usepackage{amssymb}
\usepackage{amsthm}
\usepackage{soul}

\theoremstyle{plain}% default

\newcommand{\mR}{\mathbb{R}}

\newcommand{\erf}{\operatorname{erf}}
\newcommand{\Ei}{\operatorname{Ei}}
\renewcommand{\eqref}[1]{(\ref{eq:#1})}

\begin{document}

\title{Extreme Value Laws for Superstatistics}

% Authors (Add full first names)
\author{Pau Rabassa and Christian Beck}

% Affiliations / Addresses (Add [1] after \address if there is only one affiliation.)
%\address [1] {%
\affiliation{
School of Mathematical Sciences, Queen Mary University of London,
Mile End Road, \linebreak London E1 4NS, UK}

% Contact information of the corresponding author (Add [2] after \corres if there are more than one corresponding author.)
%\corres{Email: c.beck@qmul.ac.uk }

% Abstract (Do not use inserted blank lines, i.e. \\) 
\begin{abstract}We study the extreme value distribution of stochastic processes modeled by
superstatistics. Classical extreme value theory asserts that
(under mild asymptotic independence assumptions) only three
possible limit distributions are possible, namely:  Gumbel,
Fr\'echet and Weibull  distribution. On the other hand, superstatistics
contains three important universality classes, namely $\chi^2$-superstatistics,
inverse $\chi^2$-superstatistics, and lognormal superstatistics,
all maximizing different effective entropy measures. We investigate
how the three classes of extreme value theory are related to the three classes
of superstatistics.
We show that for any superstatistical
process whose local equilibrium distribution does not live on a finite support, 
the Weibull distribution cannot occur.
Under the above mild asymptotic independence assumptions, we also show that
$\chi^2$-superstatistics generally leads an extreme value statistics
described by a Fr\'echet distribution, whereas
inverse $\chi^2$-superstatistics, as well as lognormal superstatistics,
lead to an extreme value statistics associated with the Gumbel distribution.
\end{abstract}

% Keywords: add 3 to 10 keywords
%please add severals keywrods

\maketitle

% DOCUMENT -----------------------------------------------------------------

%%%%%%%%%%%%%%%%%%%%%%%%%%%%%%%%%%%%%%%%%%%%%%%%%%%%%
\section{Introduction}
%%%%%%%%%%%%%%%%%%%%%%%%%%%%%%%%%%%%%%%%%%%%%%%%%%%%%

Superstatistics \cite{beck-cohen, swinney, touchette, jizba, chavanis, frank, celia, straeten, mark, hanel, guo, souza}
is a powerful technique to model and/or analyze complex systems
with two (or more) clearly separated time scales in the dynamics.  The basic idea is to consider
 for the theoretical modeling a superposition of many systems in local equilibrium, each with its own
inverse temperature $\beta$, and finally perform an average over the fluctuating $\beta$ which are
distributed according to
some probability density $g(\beta)$. Most generally, the parameter $\beta$
need not be inverse temperature but can be any system parameter
that exhibits large-scale fluctuations, such as energy dissipation in a turbulent
flow, or volatility in financial markets. Ultimately all expectation values relevant for the
complex system under consideration are
averaged over this distribution $g(\beta)$.
Many applications have been described in the past, including modeling the statistics of
classical turbulent flow \cite{prl2001,reynolds,swinney,prl2007},  %reference order are incorrect, pleare reorder them.
quantum turbulence~ \cite{miah},
space-time granularity \cite{jizba2}, stock price changes
\cite{straeten}, wind velocity fluctuations \cite{rapisarda},
sea level fluctuations \cite{pau1}, 
infection pathways of a virus \cite{itto},
and much more \cite{chavanis, abul-magd, soby, briggs, chen, cosmic, daniels, yalcin}.
Superstatistical systems, when integrated over the fluctuating parameter, 
are effectively described by more general entropy measures than the Boltzamnn--Gibbs entropy \cite{hanel, souza}.

In almost all of the above cases of application one will be interested in extreme values of
a suitable variable of the
complex system under consideration which is described by a particular class of superstatistics. For example, for
superstatistical models of the dynamics of share price changes, which often, in good approximation,
are either modeled by
$\chi^2$-superstatistics (equivalent to Tsallis statistics~\cite{tsallis1, tsallis-book}) or
lognormal superstatistics \cite{straeten},
extreme negative values corresponds to share price crashes. Or, for sea level fluctuations
produced by surges \cite{pau1},
extreme positive values of a surge may lead to overtopping of flood defence systems, thus leading
to flooding with all its far-reaching physical, economic, and social consequences.
For correct risk estimates of extreme events it is very important to map a given
superstatistics onto the relevant class of extreme value statistics. This is the
topic of this~paper.

Clearly, as outlined above, extreme values within a given model or data set
produced by the dynamics of a complex system are of notable practical
relevance \cite{kantz, haigh}. At the same time there is a well developed mathematical theory
for their statistical inference \cite{LLR89, EKM97, Coles01,HF06}. Recently
there has been much activity on the rigorous application of extreme values
theory to deterministic dynamical systems \cite{LFTV, FLTV, FF, FFT, FFT2,
HNT, HVR, GHN, Keller} and also to stochastically perturbed ones
\cite{AFV,FFTV,FV1}.
A remarkable feature of the dynamical system approach is that there
exist some correlations between events, and hence the extreme value
theory used to tackle it must account for this correlation going
beyond a theory that is just based on sequences of events that
are statistically independent. In the superstatistics approach,
correlations are also present, due to the fact that parameter changes
take place on long time scales, but the relaxation time of the system is short
as compared to the time scale of these parameter changes,
so that local equilibrium is quickly reached.
What is missing so far is a general analysis
which types of generalized statistical mechanics lead to which type of
extreme value statistics.
Here we deal with this question for general superstatistical
models.

Our models are general in the sense that we allow for some
mild form of statistical dependence of events; thus, it is not necessary to
have independent identically distributed random variables.
Extreme value theory quite generally tells us
(under suitable asymptotic independence assumptions) that there are only three
possible limit distributions, namely, the Gumbel,
Fr\'echet and Weibull  distribution.
On the other hand, superstatistics
contains three important universality classes, namely $\chi^2$-superstatistics,
inverse $\chi^2$-superstatistics, and lognormal superstatistics, which are quite typical
for many complex systems, meaning that most complex systems with time scale separation
fall into one of the above three classes of superstatistics.
Here we show that for any superstatistical
process whose local equilibrium distribution does not live on a finite support
the Weibull distribution cannot occur. This leaves us with Fr\'echet and Gumbel
distributions.
Under mild asymptotic independence assumptions we  show that
$\chi^2$-superstatistics generally leads to extreme values distributed according to
a Fr\'echet distribution, whereas inverse $\chi^2$ superstatistics, 
as well as lognormal superstatistics, lead to Gumbel distributions.

This paper is organized as follows.
 In Section II we briefly review the most
 important results from extreme value theory relevant for our purposes.
 Section III recalls the general concept of superstatistics,
 and provides some important results on the asymptotic behaviour of
 the generalized Boltzmann factors underlying this approach.
In Section IV we then combine the two approaches, proving our main results
which elucidate which type of extreme value theory is relevant for which type of superstatistics.
Our concluding remarks are given in Section V.

%%%%%%%%%%%%%%%%%%%%%%%%%%%%%%%%%%%%%%%%%%%%%%%%%%%%%%%%
\section{Extreme Value Theory for Stationary Processes}
\label{sec:EVT}
%%%%%%%%%%%%%%%%%%%%%%%%%%%%%%%%%%%%%%%%%%%%%%%%%%%%%%%%

Classic extreme values theory is concerned with the probability
distribution of unlikely events. Given a stationary stochastic
process $X_1,X_2, \dots$, consider the random variable
$M_n$ defined as the maximum over the first $n$ observations:
\begin{equation}
\label{eq:max process}
M_n = \max(X_1,\dots,X_n).
\end{equation}

In many cases the limit of the random variable $M_n$ may
degenerate when $n\rightarrow \infty$. Analogously to central
limit laws for partial sums, the degeneracy of the limit can be
avoided by considering a rescaled sequence $a_n(M_n - b_n)$ for
suitable normalising values $a_n\geq 0$ and $b_n\in \mR$. Indeed,
extreme value theory studies the existence of normalising values such that
\begin{equation}
\label{eq:limit-EV}
 P\left(a_n\left(M_n-b_n\right) \leq x \right) \rightarrow G\left(x\right).
\end{equation}
as $n\rightarrow \infty$, with $G(x)$ a non-degenerate probability distribution.

Two cornerstones in Extreme Value Theory are the Fisher--Tippet Theorem \cite{fisher}
and the Gnedenko Theorem \cite{gnedenko}. The former asserts that if the limiting
distribution $G$ exist, then it must be either
one of three possible types, whereas the latter theorem gives necessary
and sufficient conditions for the convergence of  each of the types.
A third cornerstone in Extreme Value Theory are the Leadbetter 
conditions \cite{Leadbetter74, LLR89}.
These are a kind of weak asymptotic independence conditions,
under which the two previous theorems generalize to stationary stochastic
series satisfying them. Let us review these results in somewhat more detail.

%--------------------------------------------------------%
\subsection{The Independent Identically Distributed Case}
%--------------------------------------------------------%

In the case where the process $X_i$ is independent identically
distributed (i.i.d.)  %Generally, we prefere to use ``iid'' instead of ``i.i.d.'', please confirm
the Fisher--Tippett Theorem states that
if $X_1, X_2, \dots$ is i.i.d. and there exist sequences
$a_n\geq 0$ and $b_n\in \mR$ such that the limit distribution
$G$ is non-degenerate, then it belongs to one of the following types:
\begin{description}
\item[Type I  :] $G(x) = \exp\left(- e^{-x} \right)$ for $x\in\mR$. This distribution is known as the
{\it Gumbel} extreme value distribution (e.v.d.).

\item[Type II :] $G(x) = \exp\left(-x^{-\alpha}\right)$, for $x>0$;
$G(x)=0$, otherwise; where $\alpha >0$ is a parameter. This family of distributions is known
as the {\it Fr\'echet} e.v.d. %Generally, we prefere to use ``evd'' or ``EVD'' instead of ``e.v.d.'', please confirm

\item[Type III: ]
$ G(x) = \exp \left(-(-x)^\alpha\right)$, for $x\leq0$; $G(x)=1$, otherwise; where
$\alpha >0$ is a parameter. This family is known as the {\it Weibull} e.v.d.
\end{description}

A further extension of this result is the Gnedenko Theorem, which provides a
characterization of the convergence in each of these cases.  Let $X_1,X_2,\dots$ 
be an i.i.d. stochastic process and let $F$ be its cumulative distribution
function. Consider $x_M=\sup\{x|\, F(x) < 1\}$.  The following conditions
are necessary and sufficient for the convergence to each type of e.v.d.:
\begin{description}
\item[Type I:]
There exists some strictly positive function $h(t)$ such that
$\lim_{t \rightarrow x_M^-} \frac{1-F(t+xh(t))}{1-F(t)} = e^{-x}$
for all real $x$;
\item[Type II:]
$x_M= +\infty$ and $\lim_{t\rightarrow \infty} \frac{1-F(tx)}{1-F(t)}= x^{-\alpha}$, with
$\alpha>0$, for each $x>0$;
\item[Type III:]
$x_M < \infty$ and $\lim_{t\rightarrow 0} \frac{1-F(x_M - tx)}{1-F(x_M - t)}= x^{\alpha}$, with
$\alpha>0$, for each $x>0$.
\end{description}

This result implies that the extremal type is completely determined
by the tail behaviour of the distribution $F(x)$.

%--------------------------------%
\subsection{The Stationary Case}
%--------------------------------%

In the case of stationary stochastic processes Leadbetter
\cite{Leadbetter74, LLR89} introduced conditions
(namely $D(u_n)$ and $D'(u_n)$) on the dependence structure which allow
for a reduction to the independent case. Given $X_1, X_2, \dots$ a stationary
sequence of random variables and $i_1,\dots,i_p$ a collection of integers, let $F_{i_1,\dots,i_p}$
denote the joint distribution function of the variables $X_{i_1},\dots,X_{i_p} $.
For brevity, we will write $F_{i_1,\dots,i_p}(u)$ for $F_{i_1,\dots,i_p}(u, \dots,u)$.
Given $\{u_n\}$ a real sequence, condition {\it $D(u_n)$}
is said to hold if for any integers $1 \leq i_1 < \dots<i_p <  j_1 < \dots<j_{p'}\leq n$
for which $j_1-i_p \geq l$, we have
\begin{equation}
\label{eq:Dun}
|F_{i_1,\dots,i_p, j_1, \dots, j_{p'}} (u_n) - F_{i_1,\dots,i_p}(u_n)
F_{j_1, \dots, j_{p'}} (u_n)| \leq \alpha_{n,l_n},
\end{equation}
where $\alpha_{n,l_n} \rightarrow 0$ as $n \rightarrow \infty$ for
some sequence $l_n = o(n)$.

Let $X_1,X_2,\dots$ be a stationary sequence and $a_n>0$ and $b_n$ given
constants such that $P\left(a_n\left(M_n-b_n\right) \leq x \right)$ converges
to a non-degenerate distribution function $G(x)$. If condition $D(u_n)$ above is
satisfied for $u_n= x/a_n + b_n$ for each real $x$, then $G(x)$ has
one of the three extreme value distribution listed before for the i.i.d. case
(see Theorem 3.3.3. in \cite{LLR89}). In other words, condition $D(u_n)$
alone is enough to extend the Fisher--Tippett Theorem to the
non-independent case.

To extend the Gnedenko Theorem (which characterizes the convergence to each
of the three extreme value types) we need to introduce an additional
condition. Given a stationary sequence $X_1,X_2,\dots$
and a sequence of constants $\{u_n\}$, condition $D'(u_n)$ will be said
to hold if
\[
\limsup_{n\rightarrow \infty} n \sum_{j=2}^{[n/k]}
P(X_1 > u_n, X_j > u_n) \rightarrow 0 \text{ as } k\rightarrow \infty,
\]
where $[ \quad]$ denotes the integer part.

Given a stationary process $X_1, X_2, \dots,$ consider the i.i.d. process
$Y_1,Y_2, \dots, $ whose distribution function is the same as that of $X_1$,
and whose partial maximum is defined as
\[
\tilde{M}_n : = \max\left(Y_1,\dots,Y_n\right) .
\] 
Suppose that $D(u_n)$ and $D'(u_n)$ are satisfied for a stationary sequence
$X_1,X_2,\dots$, when $u_n= x/a_n + b_n$ for each $x$ ($\{ a_n \}, \{b_n\}$
being given sequences of constants). Then $P\left(a_n(M_n-b_n ) \leq x \right)$ converges
to $G(x)$ for some non-degenerate $G$ d.f. if and only if
$P( a_n(\tilde{M}_n-b_n )\leq x)$ also converges to $G(x)$
(see Theorem 3.5.2. in \cite{LLR89}). In other words, if $D(u_n)$ and
$D'(u_n)$ hold, then the stochastic process $X_1, X_2, \dots$ can
be treated as if it was i.i.d.

Condition $D(u_n)$ is a weak form of mixing, which requires the stochastic
process to exhibit mild asymptotic independence of the variables, whereas
$D'(u_n)$ is a non-clustering condition. These conditions can be replaced
by stronger hypotheses, such as the $m$-dependence (requiring that $X_i$ and $X_j$
are actually independent if $|i-j| > m$) or strong mixing (a stronger
version of \eqref{Dun}). In the case of stationary normal sequences, conditions
$D(u_n)$ and $D'(u_n)$ can be replaced by requiring that the correlations
between $X_i$ and $X_j$ go to $0$ when $|i-j|$ goes to infinity.
See \cite{LLR89} for more details.

%%%%%%%%%%%%%%%%%%%%%%%%%%%%%%%%%%%%%%
\section{The Superstatistical Model}
\label{sec:SS}
%%%%%%%%%%%%%%%%%%%%%%%%%%%%%%%%%%%%%%
\vspace{-12pt}
%-------------------------------------%
\subsection{The Model}
%-------------------------------------%

Consider a non-equilibrium system that is composed of regions that
exhibit spatio-temporal fluctuations of an intensive quantity,
for example the inverse temperature $\beta$. We consider a
non-equilibrium steady state of a macroscopic system, composed of
many smaller cells that are temporarily in local equilibrium. Within
each cell, $\beta$ is approximately constant. Each cell is large enough to
obey statistical mechanics, but has a different value of the intensive
parameter $\beta$ assigned to it according to a probability density $g(\beta)$.

Given a distribution $g(\beta)$, we define
its associated {\it effective Boltzmann factor} as
\begin{equation} 
\label{eq:Boltzmann-factor} 
B(E) = \int_0^\infty g(\beta) e^{-\beta E} d\beta.
\end{equation}
The corresponding statistical mechanics based on this
generalized Boltzmann factor can be constructed if one
introduces more general entropy measures than the usual Shannon entropy
and maximizes this subject to suitable constraints,
see \cite{hanel, souza, tsallis1, kaniadakis} for examples.
If $E$ is the energy of a microstate associated with each cell, $B(E)$
represents the statistics of the statistics ($e^{-\beta E}$) of the
cells of the system. The ordinary Boltzmann factor is recovered for
$g(\beta) = \delta(\beta-\beta_0)$, where $\delta(\cdot)$ is the 
Dirac delta function.
 
As a simple paradigmatic example consider a Brownian particle that
moves in a changing environment. The particle stays for a while
in a certain cell of the system (with a given temperature), then
it moves to the next cell (with a different temperature) and so on.
In each cell the velocity $v$ obeys the equation $\dot{v} = -\gamma v
+ \sigma L(t)$, where $L(t)$ is Gaussian white noise. The inverse
temperature of each cell is related to the parameters $\gamma$
and $\sigma$ by $\beta\sim \gamma/\sigma^2$. Unlike ordinary Brownian
motion, the parameter $\beta$ is not assumed to be constant,
but fluctuates according to a probability distribution $g(\beta)$.
The stationary  probability distribution of $v$ is then
obtained by averaging the fluctuations in $\beta$.

In mathematical terms, consider a stochastic process
that for a short frame of time is well
described by a Gaussian distribution, but on a longer
time scale the parameter values of this Gaussian fluctuate. The
conditional probability density of $v$ for a given state $\beta$ is given by

 \[
f(v\,|\, \beta) \sim \exp\{-\frac{\beta v^2}{2}\}.
\]

As in the example, let $g(\beta)$ be the probability
distribution describing the fluctuations
of $\beta$, then the long-term stationary probability distribution
is obtained by averaging over $\beta$.  Therefore its
density function $f$ is
\begin{equation}
\label{eq:density-ss}
f(v) = \int_0^\infty g(\beta) f(v\,|\,\beta)d\beta \sim
 \int_0^\infty  g(\beta) e^{-\beta v^2/2} d\beta .
\end{equation}

In terms of the effective Boltzmann factor $B(\cdot)$ given by 
(\ref{eq:Boltzmann-factor}), we have that $f(v) \sim B(v^2/2)$.
The importance of the superstatistics concept comes from its generality:
One can generalize the above example to any Hamiltonian $H$ determining
the energy $E$ of the microstates, and to any distribution $g(\beta)$.
Although the superstatistics approach describes a nonequilibrium system having different regions of
different temperature, methods from equilibrium statistical mechanics
can still be formally used, such as e.g., generalized maximum entropy
principles \cite{hanel,souza}.

%-------------------------------------%
\subsection{High-Energy Asymptotics}
%-------------------------------------%

In this paper we deal with the extreme values
of a superstatistical model. To this aim, it is crucial to know the
tail behaviour of the distribution defining the process.
In the case of a superstatistical distribution, the tail
is determined by the large-$E$ behaviour of its associated
effective Boltzmann factor $B(E)$. We may use the results of \cite{touchette}
about the high-energy asymptotics.

There are three different superstatistical distributions which are
commonly found in many applications, namely $g(\beta)$ being given by a
$\chi^2$, Inverse-$\chi^2$
and Lognormal distribution. The $\chi^2$ and Inverse-$\chi^2$ superstatistics 
can be seen as representatives of a more general class of probability
distributions with a given tail behaviour: power-law tail or exponential
tail, respectively. More precisely, the exponential tail
for inverse $\chi^2$-superstatistics is an exponential
in the square root of the energy.

\subsubsection{Power-Law Tail}

Assume that the function $g(\beta)$ is such that $g(\beta) \sim \beta^\gamma$,
$\gamma>0$ as $\beta\rightarrow 0$. In this case we have that the
high energy asymptotic behaviour of the Boltzmann factor is \cite {touchette}
\[
B(E) \sim E^{-\gamma-1}, \text{ as } E\rightarrow \infty.
\]
A typical example which corresponds to the above case is that of $\beta$ being $\chi^2$-distributed.
Note that quite generally the $\beta \to 0$ behaviour of $g(\beta)$ determines
the $E \to \infty$ behavior of $B(E)$.

\subsubsection{Exponential Tail}

Assume $g(\beta)$ is such that $g(\beta) \sim e^{-c/\beta}$,
$c> 0$ as $\beta\rightarrow 0$, then the asymptotic behaviour is given by
\[
B(E) \sim E^{-3/4} e^{-2\sqrt{cE}}, \text{ as } E\rightarrow \infty.
\]
This case is realized if, for example, $g$ is equal to the inverse $\chi^2$ distribution.

\subsubsection{Log-Normal Distribution}

Assume that $g(\beta)$ is equal to the Lognormal distribution whose density 
function is 
\[
g(\beta)= \frac{1}{\sqrt{2\pi} \sigma  \beta}
\exp\left(\frac{-\left( \ln \beta - \mu \right)^2 }{2\sigma^2} \right),
\]
where $\mu$ and $\sigma>0$ are parameters.

For this example the generalized Boltzmann factor takes on the form
\[
B(E) =  \frac{1}{\sqrt{2\pi} \sigma}
\int_0^\infty
\exp\left(-\frac{\left( \ln \beta - \mu \right)^2 }{2\sigma^2}
-\beta E \right)
\frac{d\beta}{\beta}.
\]
Doing a change of variables $y= \ln \beta$ this transforms into
\[
B(E) =  \frac{1}{\sqrt{2\pi} \sigma}
\int_{-\infty}^\infty
\exp\left(-\frac{\left( y - \mu \right)^2 }{2\sigma^2} - E e^y \right)
dy.
\]

The asymptotic behaviour of $B(E)$ derived in \cite{touchette} is expressed in
terms of the Lambert or product-log function. Here we will need
more manageable asymptotics based on the asymptotics of the
characteristic function of the Lognormal distribution \cite{holgate}.
Consider

\[
\phi(t) : = \frac{1}{\sqrt{2\pi} \sigma}
\int_{-\infty}^\infty  \exp\left(-\frac{x^2 }{2\sigma^2} + i t e^{x} \right) dx.
\]
It is easy to check that
\[
B(E) = \phi(e^\mu iE).
\]
The function $\phi(t)$ corresponds to the characteristic function
of the Lognormal distribution. In \cite{holgate} its asymptotics is
studied and it is shown that

\[
\phi(t) \cong e^{it} \frac{1}{\sqrt{1-i \sigma^2 t}}
\exp\left( - \frac{\sigma^2 t^2}{2(1-i\sigma^2 t)}\right).
\]
Using this it is easy to see that
\[
B(E) \cong  \frac{1}{\sqrt{1 + \sigma^2 e^\mu E}} \exp\left(
- \frac{\frac{1}{2} \sigma^2 e^{2\mu} E^2+ e^\mu E}{\sigma^2 e^\mu E+ 1}\right),
\]
therefore
\[
B(E) \sim  E^{-1/2}  \exp{\left(- \frac{e^\mu}{2} E \right)} \]
 as $E \rightarrow \infty$.
%%%%%%%%%%%%%%%%%%%%%%%%%%%%%%%%%%%%%%%%%%%%%%%%%%%%%%%%%%%%%%
\section{Extreme Values for Superstatistical Distributions}
%%%%%%%%%%%%%%%%%%%%%%%%%%%%%%%%%%%%%%%%%%%%%%%%%%%%%%%%%%%%%%

Consider $X_1,X_2, \dots$, a stationary stochastic process for which  the
probability distribution of each random variable is well described
by a superstatistical distribution, {\em i.e.}, whose density function
$f(\cdot)$ is given by \eqref{density-ss}. We are interested in its associated
maximum process $M_n$ given by \eqref{max process} and
the existence of a limiting distribution $G$ like \eqref{limit-EV} for
suitable constants $a_n\geq 0$ and $b_n$. Additionally we will assume that
$X_1,X_2, \dots$ satisfies conditions $D(u_n)$ and $D'(u_n)$.

The Gnedenko Theorem implies that the extremal type of the limit $G$
is determined by the tail of the distribution defining the process.
When this distribution is a superstatistical distribution, its tail
is determined in turn by its associated effective Boltzmann factor.
Note that for distributions that
do not live on a finite support $f(v) > 0$ for any $v\in \mathbb{R}$, therefore
$x_M=\sup\{x|\, F(x) < 1\} = \infty$. Then, as a consequence of
Gnedenko Theorem (see Section~\ref{sec:EVT}), we have that 
convergence to a Type III (Weibull) distribution cannot occur. 

There are three different distribution functions $g$ for superstatistical
models that are commonly encountered in applications: $\chi^2$, inverse-$\chi^2$
and Lognormal. The $\chi^2$ and the inverse-$\chi^2$ superstatistics can be
understood each as particular cases of more general behaviour
of the tail, namely power-law tail and exponential law.
In this section we show that in the case of
power-law tail the associated maximum process converges to a Type II (Fr\'echet)
distribution, whereas the associated maximum process for the exponential law
and the superstatistics generated by the Lognormal distribution converges to a Type I (Gumbel) distribution.

%-------------------------------------%
\subsection{Power-Law Tail}
%-------------------------------------%

Assume that the function $g(\beta)$ is such that $g(\beta) \sim \beta^\gamma$,
$\gamma>0$ as $\beta\rightarrow 0$. In this case
the high energy asymptotic behaviour of its associated Boltzmann factor is
\[
B(E) \sim E^{-\gamma-1}, \text{ as } E\rightarrow \infty.
\]

Then, the hazard function $\bar{F}(v) = 1-F(v) = P(X_i > v)$ is given as
\[
1-F(v)  = \int_v^\infty f(u) du   = \int_v^\infty B\left(\frac{1}{2} u^2 \right) du.
\]

Given two functions $a(x)$ and $b(x)$ defined in $[x_0, \infty)$ such that
$A(x) := \int_x^\infty a(v) dv$ and $B(x) := \int_x^\infty b(v) dv$ exist and
$B(x) >0$ for any $x\in[x_0, \infty)$. Then it is straightforward (using
L'H\^opital's rule) that $a(x) \sim b(x)$ implies $A(x) \sim B(x)$.
Therefore, in the case $g(\beta) \sim \beta^\gamma$ we have
\[
1-F(v)  \sim \int_v^\infty \left(\frac{1}{2} u^2 \right)^{-\gamma-1} du
= \frac{2^{\gamma+1}}{2\gamma+1} v^{-2\gamma-1}.
\]

Using this asymptotics it is easy to check that
\[
\lim_{v\rightarrow \infty} \frac{1-F(xv)}{1-F(v)} = x^{-(2\gamma+1)}.
\]

Applying now the Gnedenko Theorem it follows that there exist
renormalizing sequences $a_n>0$ and $b_n$ such that the limiting
function $G$ associated with the maximum process $M_n$ exists and is
equal to a Fr\'echet  distribution (Type II) with shape parameter $\alpha= 2\gamma+1$.

A particular example of this case is realized when $g(\beta)$ is equal to
the $\Gamma$-distribution:
\[
g(\beta)= \frac{1}{\Gamma\left(\frac{n}{2}\right)}
\left(\frac{n}{2\beta_0} \right)^{n/2} \beta^{n/2 -1} e^{n\beta/2\beta_0},
\]
where $\beta_0$ is the average of $\beta$ and $n$ represents the
number of degrees of freedom. When $n$ is an integer and
$\beta_0=n/4$ this corresponds to a $\chi^2$ distribution with
$n$ degrees of freedom.
This distribution behaves as $g(\beta)\sim \beta^{n/2-1}$ around $\beta=0$,
which implies that the limiting function $G$ associated to its maximum
process converges to a Fr\'echet distribution with shape parameter
$\alpha= n - 1$.

%--------------------------------------------------%
\subsection{Exponential Tail}
%--------------------------------------------------%

Assume now that $g(\beta)$ is such that $g(\beta) \sim e^{-c/\beta}$,
$c> 0$ as $\beta\rightarrow 0$. In this case, its Boltzmann factor has
the following asymptotic behaviour
\[
B(E) \sim E^{-3/4} e^{2\sqrt{cE}}, \text{ as } E\rightarrow \infty.
\]

Then, the hazard function $\bar{F}(v) = 1-F(v) = P(X_i > v)$ is given as
\begin{align*}
1-F(v)  & = \int_v^\infty f(u) du   = \int_v^\infty B\left(\frac{1}{2} u^2 \right) du \\
& \sim \int_v^\infty \left(\frac{1}{2}\right)^{-3/4} u^{-3/2} e^{-2 \sqrt{c/2} u} du .
\end{align*}
as $v\rightarrow \infty$.

Using the fact that $\int v^{-3/2} e^{-v} dv = -2 \sqrt{\pi}
\erf(\sqrt{v})-(2 e^{-v})/\sqrt{v}+c$, where $\erf(\cdot)$ is the error function and
$c$ is a constant, it follows
\begin{equation}
\label{eq:hazard-exp}
1-F(v) \sim e^{-v}\left(-\left( \frac{1}{v}\right)^{3/2} +
\frac{3}{2}\left( \frac{1}{v}\right)^{5/2} + \dots \right),
\end{equation}
as $v\rightarrow \infty$.

If in this case one tries to compute the limit of $\left(1-F(xv)\right)/
\left(1-F(v)\right)$ as $v\rightarrow \infty$ one gets that
it is equal to  the limit of $e^{-(x-1)v}$ as $v\rightarrow \infty$,
which goes to $0$ or $\infty$ as $v\rightarrow \infty$.
In this case the Fr\'echet type is not a good candidate.

Using the asymptotics \eqref{hazard-exp} it is easy to check that
\[
\lim_{v\rightarrow \infty} \frac{1-F(v+x)}{1-F(v)} = e^{-x}.
\]

Applying the Gnedenko Theorem (with $h(t)\equiv 1$)
it follows that there exist renormalizing sequences
$a_n>0$ and $b_n$ such that the limiting
function $G$ associated with the maximum process $M_n$ exists and it is
equal to a Type I (Gumbel) distribution.

A particular example that falls into this class is the case where $g(\beta)$
is given by the inverse $\Gamma$~distribution
\[
g(\beta)= \frac{\beta_0}{\Gamma\left(\frac{n}{2}\right)}
\left(\frac{n\beta_0}{2} \right)^{n/2} \beta^{-n/2 -2} e^{n\beta_0/2\beta},
\]
where $\beta_0$ is again the average of $\beta$ and $n$ represents the degrees of freedom
of the inverse $\chi^2$ distribution. When $n$ is an integer and
$\beta_0=n/4$ this corresponds to an inverse $\chi^2$ distribution
with $n$ degrees of~freedom.

%-------------------------------------%
\subsection{Log-Normal Distribution}
%-------------------------------------%

Finally let us assume that $g(\beta)$ is equal to the log-normal distribution
\[
g(\beta)= \frac{1}{\sqrt{2\pi} \sigma  \beta}
\exp\left(\frac{-\left( \ln \beta - \mu \right)^2 }{2\sigma^2} \right),
\]
where $\mu$ and $\sigma>0$ are parameters. As shown in Section~\ref{sec:SS}
we have
\[
B(E) \sim  E^{-1/2}  \exp{\left(- \frac{e^\mu}{2} E \right)} \]
 as $E \rightarrow \infty$.
%{\bf (Pau: Problem as stated before)}
%
Then, the asymptotics of the hazard function $\bar{F}(v) = 1-F(v) = P(X_i > v)$
is given as
\begin{align*}
1-F(v)  & = \int_v^\infty f(u) du   =
\int_v^\infty B\left(\frac{1}{2} u^2 \right) du \\
& \sim \int_v^\infty \frac{e^{-e^\mu u^2/4}}{ u }  du .
\end{align*}
as $v\rightarrow \infty$.

Using the fact that $\int \frac{\exp(-e^u v^2/4)}{v} dv =
(\Ei(-\frac{1}{4} e^u v^2))/2+c$, where $\Ei(\cdot)$ is the exponential integral function and
$c$ is a constant, it is not difficult to check the following asymptotic expansion
\[
1-F(v) \sim
e^{-\frac{1}{4} e^\mu v^2} \left(-\frac{8e^{-\mu}}{ v^2}+O(v^{-4}) \right)
\]
as $v\rightarrow \infty$.

Using this asymptotics one can easily verify that
\[
\lim_{v\rightarrow \infty} \frac{1-F(v+xh(v))}{1-F(v)} = e^{-x},
\]
for $h(v) = \frac{2 e^{-\mu}}{v}$.

Hence the relevant extreme value distribution is again a Type I (Gumbel) distribution.

%%%%%%%%%%%%%%%%%%%%%%%%%%%%%%%%%%%%%%%%%%%%%%%%%%%%%%%%%%%%%%
\section{Conclusions and Outlook}
%%%%%%%%%%%%%%%%%%%%%%%%%%%%%%%%%%%%%%%%%%%%%%%%%%%%%%%%%%%%%%

We have considered stationary stochastic processes whose stationary
distribution functions are given in terms of
generalized Boltzmann factors, resulting from the fact
that the underlying complex system exhibits time scale
separation and is well described by a superstatistical dynamics. We have
shown that, under mild asymptotic independence assumptions,
the maximum process associated
with these superstatistical systems converges either to a Type I or a Type II
extreme value distribution. More specifically, in case of a $\chi^2$-superstatistics
(giving rise to a
Tsallis-statistics) the limiting function is equal to a Type II (Fr\'echet) distribution.
On the other hand, in the case of an inverse $\chi^2$ or a log-normal superstatistics,
the limiting function is equal to a Type I (Gumbel) distribution.

These results are relevant if one considers, for example, historical time
series of finite length, trying to extract
from the limited data set the statistics of
extreme events. Usually not enough extreme events
are available to get a reliable result. The method proposed here
would be to first analyse which type of superstatistics is realized,
and to then conclude onto the relevant extreme value distribution from our
general result derived in this paper. So for example for
temporal changes in sea levels measured at various coastal locations
in the UK (surges as produced
by meteorological changes), after subtracting the mean sea level
and the astronomical tide, a recent analysis \cite{pau1} shows that
the dynamics is
best described by a $\chi^2$-superstatistics.
Our result derived here then implies that the extreme
value statistics relevant for these sea level changes is given by 
a Fr\'echet distribution. On the other hand, measured accelerations of
a tracer particle in fully developed turbulent flows
have been shown to be well-described by log-normal superstatistics
\cite{prl2007}. Our results derived here then imply that extreme acceleration
events of this particle follow a Gumbel distribution.

\acknowledgments {Acknowledgments}

This research was supported by the EPSRC grant ``Flood MEMORY''.
We would like to thank Ivan Haigh, Matthew Wadey, Chris Kilsby, and Francesco
Serinaldi for valuable discussions.

%\authorcontributions {Author Contributions}

%Both authors have worked on this manuscript together and both authors have 
%read and approved the final manuscript.

%\conflictofinterests {Conflicts of Interest}

%The authors declare no conflict of interest. 

%%%%%%%%%%%%%%%% BIBLIOGRAPHY %%%%%%%%%%%%%%%%%%%%%%%%%%%

\end{document}